\documentclass[aps,prl,amsmath,preprintnumbers,twocolumn]{revtex4}

\usepackage{epsfig}
\usepackage{graphics}
\usepackage{latexsym}
\usepackage{amsmath}
\usepackage{amssymb}
\usepackage{rotating}
\usepackage{subfigure}
\usepackage{bm}
\usepackage{color}
\usepackage{amsfonts}
\usepackage{amsmath}
\usepackage{mathtools}

\usepackage{enumitem}

\begin{document}

ADP-15-48/T950

\title{The spin of the proton in chiral effective field theory}
\author{Hongna Li$^{1,2}$}
\author{P. Wang$^{2,3}$}
\author{D. B. Leinweber$^{4}$}
\author{A. W. Thomas$^{4,5}$}

\affiliation{$^1$College of Physics, Jilin University, Changchun, Jilin 130012, China}
\affiliation{$^2$Institute of High Energy Physics, CAS, Beijing 100049, China}
\affiliation{$^3$Theoretical Physics Center for Science Facilities, CAS, Beijing 100049,
China}
\affiliation{ $^4$Special Research Center for the Subatomic Structure
  of Matter (CSSM), Department of Physics, University of
  Adelaide, SA 5005, Australia}
\affiliation{ $^5$ ARC Centre of Excellence in Particle Physics
at the Terascale,
School of Physical Sciences, University of Adelaide, SA 5005, Australia}

\begin{abstract}
Proton spin is investigated in chiral effective field theory through
an examination of the singlet axial charge, $a_0$, and the two non-singlet
axial charges, $a_3$ and $a_8$.  Finite-range regularization is
considered as it provides an effective model for estimating the role
of disconnected sea-quark loop contributions to baryon observables.
Baryon octet and decuplet intermediate states are included to enrich
the spin and flavour structure of the nucleon, redistributing spin
under the constraints of chiral symmetry.  In this context, the proton
spin puzzle is well understood with the calculation describing all
three of the axial charges reasonably well.
The strange quark contribution to the
proton spin is negative with magnitude 0.01.  With appropriate $Q^2$
evolution, we find the singlet axial charge at the experimental scale
to be ${\hat a}_0 = 0.31^{+0.04}_{-0.05}$, consistent with the
range of current experimental values.
\end{abstract}
\maketitle


In 1988 the European Muon Collaboration (EMC) published their
polarized deep inelastic measurement of the proton's spin dependent
structure function $g_{1}$.  Their result suggested that the quark
spins summed over the up, down and strange quark flavors contribute
only a small fraction of the proton's spin~\cite{EMC}. The EMC data
shocked the particle physics community, because it was thought to be
contradictory to the apparently successful, naive quark model
descriptions of proton structure where the constituent quarks carry
the total proton spin. It inspired a vigorous global program of
experimental and theoretical developments to understand the internal
spin structure of the proton extending for nearly three decades.  For
reviews of the spin structure of the proton, see for
example Refs.~\cite{Aidala:2012mv,Myhrer:2009uq,Jaffe:2001dm,Bass:2004xa,%
  Burkardt:2008jw,Ellis:1995de,Shore:1998dn,Bass:1992bk,Filippone:2001ux}.

The experimental efforts at
CERN~\cite{Ashman:1987hv,Adeva:1998vw,Ageev:2005gh,Alexakhin:2006oza},
DESY ~\cite{Airapetian:2007mh}, JLab ~\cite{Dharmawardane:2006zd},
RHIC ~\cite{Abelev:2006uq,Adare:2007dg} and SLAC ~\cite{Hughes:1990sf}
have been impressive.  A summary of the status and recent experimental
results on the spin structure of the nucleon can be found in
Ref.~\cite{Aidala:2012mv}.  Unlike the early EMC result which suggested
that the quark
spin contribution, $\Sigma$, might be consistent with zero ($14\pm 9 \pm 21
\%$~\cite{EMC}), today the experimental measurements indicate the
nucleon's flavor-singlet axial charge measured in polarized deep
inelastic scattering is $0.35 \pm 0.03(stat.) \pm 0.05(syst.)$ at
$Q^{2}=3$ GeV$^{2}$. This tends to about one-third of the total spin $0.33 \pm
0.03(stat.) \pm 0.05(syst.)$ as $Q^{2}\rightarrow \infty$
\cite{Airapetian:2007mh,Alexakhin:2006oza,Alekseev:2010hc}.

The matrix elements of the non-singlet axial current $J_{5\mu }^{k}$ and
the singlet axial current $J_{5\mu }$ are defined as follows
\begin{eqnarray}
\left\langle p,s\right\vert \overline{\psi }\gamma ^{\mu }\gamma _{5}\frac{%
\lambda ^{k}}{2} \psi \left\vert p,s\right\rangle  &=&Ms^{\mu
}a_k,\ k=1,\, 2,\, \cdots\, 8 \, ,\quad \\
\left\langle p,s\right\vert \overline{\psi }\gamma ^{\mu }\gamma _{5} \psi\left\vert p,s\right\rangle  &=&2Ms^{\mu }a_0
=2Ms^\mu\Sigma \, ,
\end{eqnarray}
where $\lambda^k$ are generators of the flavor group and $\psi = (u,
d, s, ...)$ is a vector in flavor space.  The singlet axial current is
not conserved due to the Adler-Bell-Jackiw anomaly.  As a result, the
flavor-singlet matrix element can receive an additional contribution
from gluon
polarization~\cite{Carlitz:1988ab,Altarelli:1988nr,Bass:1991yx,Efremov:1984ip}.
This led to the early idea that the measured
singlet component $a_0$ receives an important contribution from
the gluon polarization $\Delta G$, {\it i.e.}
\begin{equation}
a_0 = \Sigma - N_f \frac{\alpha_s}{2\pi} \Delta G \, .
\end{equation}
The polarized gluon distribution function $\Delta G$ was estimated
to be less than 0.3 at a scale of 1 GeV$^2$ in the MIT bag
model~\cite{Chen:2006ng}. From the extensive experimental studies
one finds that the absolute value is
of the order $|\Delta G| \simeq$ 0.2 - 0.3 for $Q^2$ = 3
GeV$^2$~\cite{Adolph:2012ca,Nocera:2014gqa}.  This amount of gluon
polarization, by itself, is far too small to resolve the problem of
the small value of $\Sigma$ through the axial anomaly.

Another explanation for the small value of $\Sigma$ draws on the
strange quark contribution to the proton spin. The non-singlet axial
charge $a_8$ extracted from hyperon beta-decays under the assumption
of $SU(3)$ flavour symmetry is $a_8 = \Delta u + \Delta d - 2 \Delta s
= 0.58 \pm 0.03$ ~\cite{Close:1993mv}.  If the strange
quark contribution to the proton spin were around $-0.08$, the proton
spin, $\Sigma$, expressed as $a_8 + 3\Delta s$ $\simeq -0.34$,
would be close to the experimental data.  However, the uncertainty of
$a_8$ could be as large as $20 \%$
\cite{Jaffe:1989jz,Ratcliffe:2004jt}.  A recent re-evaluation of the
nucleon's axial-charges in the Cloudy Bag model, taking into account
the effect of the one-gluon-exchange hyperfine interaction and the
meson cloud, led to the value $a_8 = 0.46 \pm 0.05$
\cite{Bass:2009ed}. In this case, $\Delta s$ was found to be of
order 0.01 in magnitude (and negative), with the small value of
$a_8$ a consequence of SU(3) breaking.

Soon after the release of the EMC data it was realized that the
effect of the pion cloud of the nucleon, associated with chiral
symmetry breaking, would be to lower the quark spin content of the
nucleon~\cite{Schreiber:1988uw}. This is because pion emission
tends to flip the nucleon spin and hence the spins of the quarks in
it, while the quarks in the pion necessarily carry orbital
angular momentum but no spin. This effect was calculated in the
cloudy bag model and the effect of the pion cloud together with
the relativistic
motion of the light quarks in the bag~\cite{Chodos:1974je} reduced
$\Sigma$ to around 0.5. An alternative approach to the problem recognised
that, given the standard spin dependent one-gluon-exchange correction to the
energy of the nucleon, there must be a corresponding exchange current
correction to the proton spin~\cite{Myhrer:1988ap}.
This too reduces the proton spin by
around 0.15 below the naive bag model result of 0.65.
It is only recently that studies of the $\Delta$ nucleon mass splitting
in lattice QCD~\cite{Young:2002cj} provided the justification for
combining the pion cloud
and one-gluon exchange effects~\cite{Myhrer:2009uq}.
This led to a theoretical result in the range $0.35$ to $0.40$,
which is compatible with the aforementioned
experimental value and, after QCD evolution, with the results
of lattice QCD for the angular momentum carried by the
quarks in the proton~\cite{Thomas:2008ga}.

The scale dependence of $a_0$ presents another consideration in
understanding the fraction of the proton spin carried by
quarks~\cite{Jaffe:1987sx}. Consideration of the genral features of
QCD evolution long ago led to the conclusion that the natural scale
at which to match a quark model to QCD is quite low, so that most of
the momentum of the proton is carried by valence quarks and one
can think of the gluons as having been integrated out of the theory.
In Jaffe's scenario, the small value of the
experimental proton spin is due to differences in the energy scale of
the experimental result and the quark model results.  Since the
anomalous dimension of the singlet axial current is nontrivial, its
matrix element $a_0$ is scale dependent.  With the $Q^2$
evolution, it is possible that the large proton spin at low $Q^2$ will
be reduced through $Q^2$ evolution to the large $Q^2$ of the
experimental result.  As mentioned in Ref.~\cite{Jaffe:1987sx}, it is
difficult to get a reliable evolution at low-$Q^2$ because
perturbative QCD is not applicable.  More specifically, one cannot
determine which evolution line presented in Ref.~\cite{Jaffe:1987sx}
is correct.  One needs a direct calculation of $\Sigma$ at the low
energy scale.

In this paper, we will investigate the proton spin carried by
the quarks in the framework of effective field theory,
assuming that at the corresponding low scale the gluons have been
integrated out, with the only residue being a spin-dependent
effective interaction between quarks.
$\Sigma$, and the non-singlet axial charges $a_3$ and $a_8$
will be calculated simultaneously in the chiral effective field
theory.  In this approach the proton
structure is enhanced through the dressing of the proton by
octet-meson and both octet and decuplet baryon intermediate states.
These processes enrich the spin and flavour structure of the nucleon,
redistributing spin under the constraints of chiral symmetry.  As we
will see, this formalism is able to describe all three axial charges in
a reasonable manner.

We consider heavy baryon chiral perturbation theory and include octet
and decuplet intermediate-state baryons.  The lowest-order chiral
Lagrangian used in the calculation of the nucleon spin distribution
function is expressed as
\begin{eqnarray}
{L_{v}}&=&iTr\overline{{B}}{_{v}}\left( {v\cdot {\cal D}}\right) {B_{v}}+2DTr
\overline{{B}}{_{v}}S_{v}^{\mu }\left\{ {{A_{\mu }},{B_{v}}}\right\}
\nonumber \\
&&+2FTr\overline{{B}}{_{v}}S_{v}^{\mu }\left[ {{A_{\mu }},{B_{v}}}\right]
-i\overline{T}_{v}^{\mu }\left( {v\cdot {\cal D}}\right) {T_{v\mu }}
\nonumber \\
&&+{\cal C}\left( {\overline{T}_{v}^{\mu }{A_{\mu }}{B_{v}}+\overline{{B}}_{v}{A_{\mu
}}T_{v}^{\mu }}\right) \, ,
\end{eqnarray}
where $S_{v}^{\mu }$ is the covariant spin operator defined as
\begin{equation}
S_{v}^{\mu }=\frac{i}{2}{\gamma ^{5}}{\sigma ^{\mu \nu }}{v_{\nu }}\, .
\end{equation}%
Here, $v^{\nu }$ is the nucleon four velocity.  In the rest frame, we
have $ v^{\nu }=(1,0,0,0)$). $D$, $F$ and ${\cal C}$ are the standard
$SU(3)$-flavour coupling constants.
\begin{figure}[t]
\includegraphics[width=7cm]{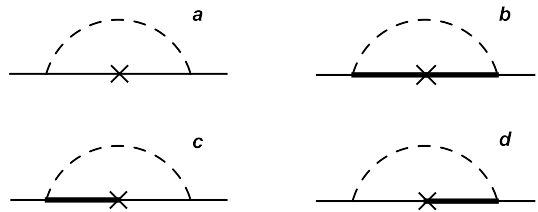}
\caption{The one-loop Feynman diagrams for calculating the quark
  contribution to the proton spin. The thin and thick solid lines are
  for the octet and decuplet baryons, respectively. }
\label{fig:1}
\end{figure}

According to the Lagrangian, the one-loop Feynman diagrams, which
contribute to the quark spin fraction of the proton, are plotted in
Fig.~1.  Working with the chiral coefficients of full
QCD~\cite{Leinweber:2002qb,Wang:2007iw}, the contribution of the
doubly-represented $u$-quark sector of the proton to the proton spin,
described by diagram (a) of Fig.~1, is expressed as
\begin{eqnarray}
\Delta u^{a}&=& \left [ C_{N\pi }\, I_{2\pi }^{NN}+C_{\Sigma K}\, I_{2K}^{N\Sigma
}+C_{\Lambda \Sigma K}\, I_{5K}^{N\Lambda \Sigma } \right . \nonumber \\
&&\left . + C_{N\eta }\, I_{2\eta }^{NN}\right ]s_{u} \, ,
\end{eqnarray}
where the first through fourth terms in the bracket are the
contributions from the $\pi N$, $K \Sigma$, the $K(\Lambda-\Sigma)$
transition and the $\eta N$ intermediate states, respectively.
The $u$-quark contribution with a $\Lambda$
intermediate state vanishes. The coefficients, $C$, of the
integrals, $I$, are expressed as

\begin{eqnarray}
C_{N\pi}
&=&-\frac{(D+F)^2}{288\,\pi^3\, f_\pi^2} \, , \\
C_{\Sigma K}
&=&-\frac{5(D-F)^{2}}{288\,\pi^3\, f_\pi^2} \, , \\
C_{\Lambda \Sigma K}
&=&\frac{(D-F)\,(D+3F)}{288\,\pi^3\, f_\pi^2} \, , \\
C_{N\eta }
&=&-\frac{2}{3}\, \frac{(3F-D)^{2}}{288\,\pi^3\, f_\pi^2} \, .
\end{eqnarray}

These coefficients reflect the $SU(3)$-flavour symmetry considered in
obtaining the meson-baryon couplings (proportional to $F$ and $D$),
the angular-momentum composition of the intermediate meson-baryon
intermediate states, and the $SU(6)$-spin-flavour wave function of the
intermediate state baryon, considered in assigning a quark-sector spin
contribution.  The latter is discussed in further detail below.

With the above coefficients, one can write the $d$ quark-sector contribution to
the proton spin of Fig.~1a as
\begin{eqnarray}
\Delta d^{a}&=& \left [ \frac{7}{2}C_{N\pi }\, I_{2\pi }^{NN} +
  \frac{1}{5} C_{\Sigma K}\, I_{2K}^{N\Sigma } - C_{\Lambda \Sigma
    K}\, I_{5K}^{N\Lambda \Sigma }
  \right . \nonumber \\
&&\left . -\frac{1}{4} C_{N\eta }\, I_{2\eta }^{NN}\right ] s_{d}.
\end{eqnarray}
Similarly, the strange quark contribution to the proton spin from
diagram (a) of Fig.~1 is written as
\begin{equation}
\Delta s^{a} = \left [ -\frac{3}{10}C_{\Sigma K}\, I^{N\Sigma}_{2K}
               + C_{\Lambda K }\, I^{N\Lambda}_{2K} \right ] s_s \, ,
\end{equation}
where

\begin{equation}
C_{\Lambda K} = -\frac{1}{2}\,\frac{(D+3F)^{2}}{288\,\pi^3\, f_\pi^2} \, .
\end{equation}

In the above equations, the low energy coefficients $s_q$ ($q= u, d,
s$) describe the tree-level quark contribution to the baryon spin. For
example, for the intermediate proton and neutron, their spins are
expressed as
\begin{equation}
s_p = \frac{4}{3} s_u - \frac{1}{3} s_d\, , ~~~ s_n = \frac{4}{3} s_d -
\frac{1}{3} s_u \, .
\end{equation}
In the naive quark model, the value of $s_q$ is 1.  However, it is
smaller than 1 due to the relativistic and confinement effects
\cite{Chodos:1974je}.

Diagram (b) of Fig.~1 illustrates decuplet baryon intermediate
states. The $u$-sector contribution to the proton spin from this
diagram is
\begin{equation}
\Delta u^{b}= \left [ C_{\Delta\pi}\, I^{N\Delta}_{2\pi} + C_{\Sigma^*
    K}\, I^{N\Sigma^*}_{2K} \right ]\, s_u \, ,
\end{equation}
where the coefficients $C_{\Delta\pi}$ and $C_{\Sigma^* K}$ are
\begin{eqnarray}
C_{\Delta\pi}  &=& \frac{35\, {\cal C}^2}{648\, \pi^3\, f_\pi^2} \, , \\
C_{\Sigma^* K} &=& \frac{5}{28}\, C_{\Delta\pi} \, .
\end{eqnarray}
The $d$ and $s$ quark-sector contributions are
\begin{equation}
\Delta d^{b}= \left [ \frac{2}{7}\, C_{\Delta\pi}\, I^{N\Delta}_{2\pi}
  + \frac{1}{5}\, C_{\Sigma^* K}\, I^{N\Sigma^*}_{2K}\right ]\, s_d,
\end{equation}
and
\begin{equation}
\Delta s^{b}= \frac{3}{5}\, C_{\Sigma^* K}\, I^{N\Sigma^*}_{2K}\, s_s.
\end{equation}
In deriving these equations, the tree-level quark contributions to the
spin of decuplet baryons are used.  For example
\begin{equation}
s_{\Delta^+} = 2\, s_u + s_d\, , ~~~ s_{\Sigma^{*-}} = 2\, s_d + s_s \, .
\end{equation}
  These contributions will also be reduced upon taking relativistic
  and confinement effects into account.

Diagrams (c) and (d) of Fig.~1 provide contributions from intermediate
states involving an octet-decuplet transition.  The $u$ quark-sector
contribution to the proton spin from these diagrams is expressed as
\begin{eqnarray}
\Delta u^{c+d} &=& \left [ C_{N\Delta\pi}\, I^{N\Delta}_{3\pi}
+ C_{\Sigma\Sigma^* K}\, I^{N\Sigma\Sigma^*}_{5K}
+ C_{\Lambda\Sigma^* K}\, I^{N\Lambda\Sigma^*}_{5K} \right ] \nonumber \\
&& \times  s_u \, ,
\end{eqnarray}
where

\begin{eqnarray}
C_{N\Delta\pi}        &=& -\frac{(D+F)\,{\cal C}}{27\, \pi^3\, f_\pi^2} \, , \\
C_{\Sigma\Sigma^* K}  &=& -\frac{5}{8}\, \frac{(D-F) \,{\cal C}}{27\, \pi^3\, f_\pi^2} \, , \\
C_{\Lambda\Sigma^* K} &=& -\frac{1}{8}\, \frac{(D+3F)\,{\cal C}}{27\, \pi^3\, f_\pi^2} \, .
\end{eqnarray}

The $d$ and $s$ quark-sector contributions are
\begin{eqnarray}
\Delta d^{c+d}&=& \left [ -C_{N\Delta \pi }\, I_{3\pi }^{N\Delta }
+\frac{1}{5}\, C_{\Sigma \Sigma ^{\ast }K}\, I_{5K}^{N\Sigma \Sigma
  ^{\ast }} \right . \nonumber \\
&&\qquad \left . -C_{\Lambda \Sigma ^{\ast }K}\, I_{5K}^{N\Lambda \Sigma ^{\ast
}}\right ]\, s_{d}\, , \\
\Delta s^{c+d}&=&-\frac{6}{5}\, C_{\Sigma \Sigma ^{\ast }K}\, I_{5K}^{N\Sigma \Sigma
^{\ast }}\, s_{s}\, .
\end{eqnarray}
The integrals in the above equations, $I^{\alpha\beta}_{2j}$,
$I^{\alpha\beta\gamma}_{5j}$ and $I^{\alpha\beta}_{3j}$ are defined in
Ref.~\cite{Wang:2007iw}.

Including the tree-level contribution, the total $u$-, $d$- and
$s$-quark sector contributions to the spin of the  proton are
\begin{eqnarray} \nonumber
\Delta {u}&=&\frac{4}{3}Z{s_{u}}+\Delta u^{a}+\Delta u^{b}+\Delta u^{c+d}\, , \\
\Delta {d}&=&-\frac{1}{3}Z s{_{d}}+\Delta d^{a}+\Delta d^{b}+\Delta d^{c+d}\, , \nonumber \\
\Delta {s}&=&\Delta s^{a}+\Delta s^{b}+\Delta s^{c+d}\, .
\end{eqnarray}
Here $Z$ is the wave-function renormalization constant calculated from
the standard diagrams corresponding to those of Fig.~1.  Values are
listed in Table I.

In the numerical calculations, the $SU(3)$-flavour couplings are
$D=0.8$, $F=0.46$. The decuplet coupling ${\cal C} =
-1.2$ \cite{Jenkins:1992pi}.  The regulator in the integrals is chosen
to be of a dipole form
\begin{equation}
u(k) =\frac{1}{\left(1+k^2/\Lambda^2\right)^2} \, ,
\end{equation}
with $\Lambda=0.8 \pm 0.2$ GeV.  This prescription is known to model the
contributions of disconnected sea-quark loop contributions
well~\cite{Leinweber:2004tc,Wang:1900ta,Wang:2012hj}.

The final quark spin contributions are related to the low-energy
coefficients $s_u$, $s_d$ and $s_s$.  These are the tree-level values
of the quark spin and are unity in the naive constituent-quark model.
Relativistic and confinement effects associated with light quarks
suppress this value.  We begin by assuming $s_u=s_d=s_s=s_q$ and treat
$s_q$ as a parameter constrained by the axial charge $a_3=1.27$.
With $\Lambda = 0.8$ GeV, the central value of $s_q$ that we
find is $s_q=0.79$, less than one as expected.  Since the strange quark
is expected to be less relativistic, this value may be an overestimate
of the spin suppression in that case.
However, the strange quark contribution to
the proton spin is small and so the approximation is adequate for this
purpose.

With $s_q=0.79$, the $u$, $d$ and $s$ quark contributions to the
proton spin are
\begin{equation}
\Delta u = 0.94\, , ~~~~ \Delta d = -0.33\, , ~~~~ \Delta s = -0.01\, .
\end{equation}
The axial charge $a_8=0.63$ and $\Sigma=0.61$.
Before considering the necessary $Q^2$ evolution to the value $Q^{2}=3$
GeV$^{2}$ relevant to the experimental data, it is interesting to consider
other improvements to our use of $SU(6)$-spin-flavour wave functions
in attributing quark spin to intermediate meson-baryon states.

Although it lies outside the framework of chiral effective field theory,
the effect of one-gluon-exchange (OGE) is particularly important for
spin dependent quantities. Hogaason and Myhrer~\cite{Hogaasen}
showed that the incorporation of
the exchange current correction arising from the effective
one-gluon-exchange (OGE) force
shifts the tree-level non-singlet charge, $a_3$, from
$\frac{5}{3} s_q$ to $\frac{5}{3} s_q - G$, where $G$ is about 0.05.
Thus, if one were to include the OGE correction,
$s_q$ would be somewhat larger at $0.82$
if one chose it to reproduce the axial charge $a_3 = 1.27$.
{}For the charges,
$a_0$ and $a_8$, the OGE correction shifts their tree-level values
from $s_q$ to $s_q-3G$~\cite{Myhrer:2009uq}.
In this case, $a_0 = 0.51$ and $a_8 =
0.53$.  Correspondingly, the quark contributions to the proton spin
are
\begin{equation}
\Delta u = 0.90\, , ~~~~ \Delta d = -0.38\, , ~~~~ \Delta s = -0.01\, .
\end{equation}
The results show that the strange quark contribution to the proton
spin is very small relative to the $u$ and $d$ contributions.  The
axial charge, $a_8 = 0.53$, is intermediate between the
value extracted under the assumption of SU(3) symmetry from hyperon
$\beta$-decay, $0.58 \pm 0.03$ \cite{Close:1993mv}, and that obtained
in the cloudy bag model, $0.46 \pm 0.05$~\cite{Bass:2009ed}.

To provide an estimate of the uncertainty in these results, we vary
the regulator parameter, $\Lambda$, governing the size of meson
cloud contributions to proton structure.  Considering $\Lambda = 0.8
\pm 0.2 $ GeV, the uncertainties in the quark contributions to proton
spin are
\begin{eqnarray}
\Delta u &=& +0.90\,^{+0.03}_{-0.04}\, , \\
\Delta d &=& -0.38\,^{+0.03}_{-0.03}\, , \\
\Delta s &=& -0.007\,^{+0.004}_{-0.007}\, .
\end{eqnarray}
The axial charges with the corresponding error bars are
\begin{equation}
 a_0=\Sigma = 0.51\,^{+0.07}_{-0.08} \, ,
\quad\mbox{and}\quad
   a_8=0.53\,^{+0.06}_{-0.06}\, .
\label{eq:a0a8final}
\end{equation}

The non-singlet axial current is conserved in the limit of massless
quarks and the anomalous dimension for the non-singlet axial current
vanishes.  Therefore, the non-singlet matrix elements $a_3$ and $a_8$
are scale independent.
However, the anomalous dimension of the singlet axial current is
nontrivial, and $a_0$ is a scale dependent quantity. Consistent with
the idea that at a sufficiently low scale the valence quarks dominate
and the gluons
have been effectively integrated out of the theory, we set
$a_0=\Sigma$ at that scale.
Then to compare the result for $a_0$ calculated within
chiral effective field theory
to experiment, $\hat{a}_0(Q^2)$ is obtained through
NNLO QCD evolution to $Q^{2}=3$ GeV$^{2}$.
\begin{figure}[t]
\includegraphics[width=8.5cm]{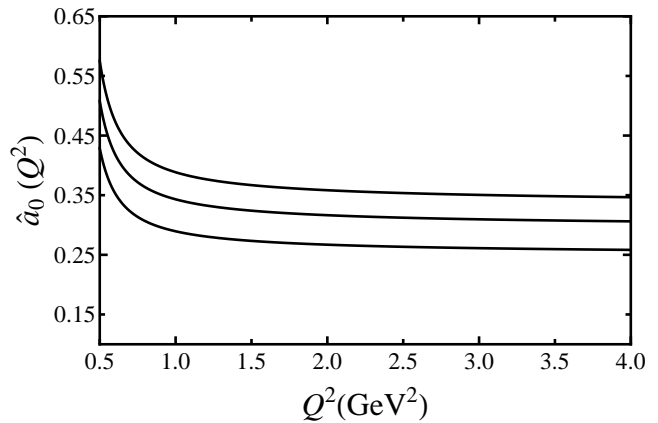}
\caption{$Q^2$ evolution of the singlet axial charge of
  Eq.~(\ref{eq:a0a8final}), $a_0 = {\hat a}_0(\mu=0.5\mbox{ GeV}^2)$
  for the proton. The upper, middle and lower lines are for the values
  $\Lambda = 0.6$, 0.8 and 1 GeV, respectively and provide insight
  into the role of the meson cloud and the sensitivity of our result
  at $\Lambda = 0.8$ to variations in the size of the meson-cloud
  dressing of the proton.}
\label{fig:2}
\end{figure}

The $Q^2$ evolution equation has the form~\cite{Gensini:1995ev}
\begin{equation}
\frac{d}{dt}\, \hat{a}_0 (t) = - N_f\,
\frac{\alpha_s}{2\pi} \, \gamma_{gq}\, \hat{a}_0(t) \, ,
\end{equation}
where $t=\log Q^2/\mu^2$.
After integrating in $\alpha_s$ from a normalization scale
of $\mu^2$ to $Q^2$, one obtains~\cite{Gensini:1995ev}
\begin{eqnarray}
\log \frac{{\hat a}_0(Q^2)}{{\hat a}_0(\mu^2)} &=&
\frac{6N_f}{33-2N_f}\, \frac{\alpha _s(Q^2)-\alpha _s(\mu^2)}{\pi}
\nonumber \\
&&\times \left [ 1 + \left( \frac{83}{24} + \frac{N_f}{36} - \frac{33-2N_f}{8(153-19N_f)} \right )
\right . \nonumber \\
&&\qquad \quad \left . \times
\frac{\alpha_s(Q^2)+\alpha _s(\mu^2)}{\pi} \right ] \, ,
\label{evolution}
\end{eqnarray}
with the NNLO calculation of the anomalous dimension, $\gamma_{gq}$,
taken from Ref.~\cite{Larin:1993tq}.

In Fig.~2 we illustrate the $Q^2$ evolution of $\hat{a}_0 (Q^2)$
commencing with our result of Eq.~(\ref{eq:a0a8final}) attributed to
the scale $\mu = 0.5$~GeV$^2$ as in Ref.~\cite{Jaffe:1987sx}.
Initially, $\hat{a}_0 (Q^2)$ decreases rapidly with increasing $Q^2$
raising concerns about the application of an NNLO calculation for $Q^2
< 1\mbox{ GeV}^2$.  However, in the context of the model uncertainty
presented in Fig.~2 the present $Q^2$ evolution will suffice.

At $Q^2=3$ GeV$^2$, our calculation of the proton spin can be compared
with experiment.  Our model provides
\begin{equation}
\hat{a}_0 (3\mbox{ GeV}^2) = 0.31\,^{+0.04}_{-0.05}\, ,
\end{equation}
which agrees with the experimental measurement of $0.35 \pm 0.03(stat.)
\pm 0.05(syst.)$ at $Q^{2}=3$ GeV$^{2}$.
\begin{table*}[tbp]
\caption{The predictions of the meson-cloud model presented herein for
  proton spin structure as a function of the regulator parameter,
  $\Lambda = 0.8 \pm 0.2$, governing the size of the meson-cloud
  dressings of the proton.}
\label{tab1}
\begin{ruledtabular}
\begin{tabular}{cccccccccc}
$\Lambda$ (GeV)& $Z$ & $s_q$ & $\Delta u$ & $\Delta d$ & $\Delta s$
& $g_A$ & $a_8$ & $\Sigma$ & ${\hat a}_0$ (3 GeV$^2$)  \\ \hline
0.6 & 0.84 & 0.83 & 0.93 & $-0.35$ & $-0.003$ &  1.27  & 0.59 & 0.58 & 0.35 \\
0.8 & 0.71 & 0.82 & 0.90 & $-0.38$ & $-0.007$ & 1.27 & 0.53 & 0.51 & 0.31 \\
1.0 & 0.58 & 0.76& 0.86 & $-0.41$ & $-0.014$ & 1.27 & 0.47 & 0.43 & 0.26 \\
\end{tabular}
\end{ruledtabular}
\end{table*}

In summary, we have examined the proton spin fractions carried by
quarks using a model in which the meson-cloud dressings of the proton
are characterized by chiral effective field theory, regularized through
a regulator characterizing the nontrivial size of the source of the
meson cloud.
Finite-range regularization provides an effective model for estimating
the role of disconnected sea-quark loop contributions to baryon
observables
\cite{Leinweber:2004tc,Wang:1900ta,Wang:2012hj,Hall:2013dva,Wang:2013cfp,Wang:2015sdp}.
Both baryon octet and decuplet intermediate states are included to
enrich the spin and flavour structure of the nucleon, redistributing
spin under the constraints of chiral symmetry.
Drawing on extensive experience
\cite{Leinweber:2004tc,Wang:1900ta,Wang:2013cfp,Wang:2012hj,Young:2002ib,%
Leinweber:2003dg,Wang:2007iw,Wang:2010hp,Allton:2005fb,Armour:2008ke,Hall:2013oga},
the preferred regulator parameter is $\Lambda = 0.8$ GeV.
To gain insight into the role of the meson cloud and uncertainties
associated in determining the size of the meson-cloud contributions,
we have varied $\Lambda$ from 0.6 GeV to 1 GeV.

The coefficient $s_q$, which takes relativistic and confinement
effects into account is constrained by the experimental axial charge
$a_3=g_A=1.27$.
The one-gluon-exchange correction to the axial charges is also taken
into consideration.
Because each quark-sector contribution is calculated separately,
the non-singlet charges, $a_3$ and $a_8$, and the singlet charge $a_0$ are
obtained simultaneously.  The results are summarized in Table I.

Our model provides significant insight into the the proton spin puzzle.
The main conclusions are:
\begin{enumerate}[leftmargin=0cm,itemindent=.5cm,labelwidth=\itemindent,labelsep=0cm,align=left]
\item
At low energy scales the total quark spin contribution to the proton spin,
$\Sigma=0.51^{+0.07}_{-0.08}$, is only of order one half in the valence region.

\item
As indicated in Table \ref{tab1}, all three of the quark spin
contributions $\Delta u$, $\Delta d$ and $\Delta s$ decrease in value
as one increases the size of the meson-cloud contribution by
increasing $\Lambda$.  As a result the net spin
carried by the quarks, $\Sigma$,
diminishes with increasing meson-cloud contributions.  This is in
accord with the increased role of orbital angular
momentum~\cite{Thomas:2008ga} between the
odd-parity mesons and the even-parity baryons of the proton's meson
cloud considered herein.

\item
The parameter $s_q$ reflecting the role of relativistic and
confinement effects and constrained by $a_3$ is around 0.82, smaller than 1
as expected but larger than the typical ``ultra-relativistic'' value
of 0.65.  Again, increasing the size of the meson-cloud contributions
diminishes this value.  For example, at $\Lambda = 1$ GeV, $s_q =
0.76$.

\item
The non-singlet charge $a_8=0.53\,^{+0.06}_{-0.06}$
lies between the
value extracted from the hyperon $\beta$-decays
under the assumption of SU(3) symmetry, $0.58\pm 0.03$, and
the value $0.46 \pm 0.05$ obtained in the cloudy bag model~\cite{Bass:2009ed}.
Because the experimental value of $a_0$ extracted from DIS data depends
on this quantity, further work to pin down the extent of SU(3)
breaking would be valuable.

\item
The strange quark contribution to the proton spin is negative and its
absolute value is of the order 0.01.  Larger $\Lambda$ values admit
stronger hyperon contributions which act to increase this magnitude.

\item
The experimental value of the $a_0$ at 3 GeV$^2$ is reproduced
through a combination of the chiral correction and
$Q^2$ evolution of $\Sigma$ from a scale of 0.5 GeV$^2$~\cite{Jaffe:1987sx}.
We find
$\hat{a}_0$ (3 GeV$^2$) is $0.31^{+0.04}_{-0.05}$ which agrees with
the experimental measurement of $0.35 \pm 0.03(stat.)  \pm
0.05(syst.)$.

\end{enumerate}

Future work should explore the role of higher order terms in the $Q^2$
evolution of $a_0$ and explore non-perturbative treatments that can
provide further insight into the connection between models of hadron
structure and modern experimental results.

\section*{Acknowledgments}

This work is supported in part by DFG and NSFC (CRC 110), by the
National Natural Science Foundation of China (Grant No. 11475186) and
by the Australian Research Council through grants DP140103067,
DP150103164 (DBL), FL0992247 and DP151103101 (AWT) and through the ARC Centre of
Excellence for Particle Physics at the Terascale.


\begin{thebibliography}{99}

\bibitem{EMC}  J.~Ashman \textit{et al.} [European Muon Collaboration],
Phys.\ Lett.\ B \textbf{206}, 364 (1988).  

\bibitem{Aidala:2012mv}  C.~A.~Aidala, S.~D.~Bass, D.~Hasch and
G.~K.~Mallot,  
Rev.\ Mod.\ Phys.\ \textbf{85}, 655 (2013)  [arXiv:1209.2803 [hep-ph]].

\bibitem{Myhrer:2009uq}  F.~Myhrer and A.~W.~Thomas,
J.\ Phys.\ G \textbf{37}, 023101 (2010)  [arXiv:0911.1974 [hep-ph]].

\bibitem{Jaffe:2001dm}  R.~L.~Jaffe,
AIP Conf.\ Proc.\ \textbf{588}, 54 (2001)  [hep-ph/0102281].

\bibitem{Bass:2004xa}  S.~D.~Bass,  
Rev.\ Mod.\ Phys.\ \textbf{77}, 1257 (2005)  [hep-ph/0411005].

\bibitem{Burkardt:2008jw}  M.~Burkardt, C.~A.~Miller and W.~D.~Nowak,
Rept.\ Prog.\ Phys.\ \textbf{73}, 016201 (2010)  [arXiv:0812.2208 [hep-ph]].

\bibitem{Ellis:1995de}  J.~R.~Ellis and M.~Karliner,
[hep-ph/9601280].  

\bibitem{Shore:1998dn}  G.~M.~Shore,
[hep-ph/9812355].  

\bibitem{Bass:1992bk}
  S.~D.~Bass and A.~W.~Thomas,
  J.\ Phys.\ G {\bf 19}, 925 (1993).

\bibitem{Filippone:2001ux}  B.~W.~Filippone and X.~D.~Ji,
Adv.\ Nucl.\ Phys.\ \textbf{26}, 1 (2001)  [hep-ph/0101224].

\bibitem{Ashman:1987hv}
  J.~Ashman {\it et al.}  [European Muon Collaboration],
  Phys.\ Lett.\ B {\bf 206}, 364 (1988).

\bibitem{Adeva:1998vw}
  B.~Adeva {\it et al.}  [Spin Muon Collaboration],
  Phys.\ Rev.\ D {\bf 58}, 112002 (1998).

\bibitem{Ageev:2005gh}
  E.~S.~Ageev {\it et al.}  [COMPASS Collaboration],
  Phys.\ Lett.\ B {\bf 612}, 154 (2005)
  [hep-ex/0501073].

\bibitem{Alexakhin:2006oza}
  V.~Y.~Alexakhin {\it et al.}  [COMPASS Collaboration],
  Phys.\ Lett.\ B {\bf 647}, 8 (2007)
  [hep-ex/0609038].

\bibitem{Airapetian:2007mh}
  A.~Airapetian {\it et al.}  [HERMES Collaboration],
  Phys.\ Rev.\ D {\bf 75}, 012007 (2007)
  [hep-ex/0609039].

\bibitem{Dharmawardane:2006zd}
  K.~V.~Dharmawardane {\it et al.}  [CLAS Collaboration],
  Phys.\ Lett.\ B {\bf 641}, 11 (2006)
  [nucl-ex/0605028].

\bibitem{Abelev:2006uq}
  B.~I.~Abelev {\it et al.}  [STAR Collaboration],
  Phys.\ Rev.\ Lett.\  {\bf 97}, 252001 (2006)
  [hep-ex/0608030].

\bibitem{Adare:2007dg}
  A.~Adare {\it et al.}  [PHENIX Collaboration],
  Phys.\ Rev.\ D {\bf 76}, 051106 (2007)
  [arXiv:0704.3599 [hep-ex]].

\bibitem{Hughes:1990sf}
  V.~W.~Hughes,
  Nucl.\ Phys.\ A {\bf 518}, 371 (1990)£¬and references therein.

\bibitem{Alekseev:2010hc}
  M.~G.~Alekseev {\it et al.}  [COMPASS Collaboration],
  Phys.\ Lett.\ B {\bf 690}, 466 (2010)
  [arXiv:1001.4654 [hep-ex]].

\bibitem{Carlitz:1988ab}
  R.~D.~Carlitz, J.~C.~Collins and A.~H.~Mueller,
  Phys.\ Lett.\ B {\bf 214}, 229 (1988).

\bibitem{Altarelli:1988nr}
  G.~Altarelli and G.~G.~Ross,
  Phys.\ Lett.\ B {\bf 212}, 391 (1988).

\bibitem{Bass:1991yx}
  S.~D.~Bass, B.~L.~Ioffe, N.~N.~Nikolaev and A.~W.~Thomas,
  J.\ Moscow.\ Phys.\ Soc.\  {\bf 1}, 317 (1991).

\bibitem{Efremov:1984ip}
  A.~V.~Efremov and O.~V.~Teryaev,
  Phys.\ Lett.\ B {\bf 150}, 383 (1985).

\bibitem{Chen:2006ng}
  P.~Chen and X.~Ji,
  Phys.\ Lett.\ B {\bf 660}, 193 (2008)
  [hep-ph/0612174].

\bibitem{Adolph:2012ca}
  C.~Adolph {\it et al.}  [COMPASS Collaboration],
  Phys.\ Rev.\ D {\bf 87}, no. 5, 052018 (2013)
  [arXiv:1211.6849 [hep-ex]].

\bibitem{Nocera:2014gqa}
  E.~R.~Nocera {\it et al.}  [NNPDF Collaboration],
  Nucl.\ Phys.\ B {\bf 887}, 276 (2014)
  [arXiv:1406.5539 [hep-ph]].

\bibitem{Close:1993mv}
  F.~E.~Close and R.~G.~Roberts,
  Phys.\ Lett.\ B {\bf 316}, 165 (1993)
  [hep-ph/9306289].

\bibitem{Jaffe:1989jz}
  R.~L.~Jaffe and A.~Manohar,
  Nucl.\ Phys.\ B {\bf 337}, 509 (1990).

\bibitem{Ratcliffe:2004jt}
  P.~G.~Ratcliffe,
  Czech.\ J.\ Phys.\  {\bf 54}, B11 (2004)
  [hep-ph/0402063].

\bibitem{Bass:2009ed}
  S.~D.~Bass and A.~W.~Thomas,
  Phys.\ Lett.\ B {\bf 684}, 216 (2010)
  [arXiv:0912.1765 [hep-ph]].

\bibitem{Schreiber:1988uw}
  A.~W.~Schreiber and A.~W.~Thomas,
  Phys.\ Lett.\ B {\bf 215}, 141 (1988).

\bibitem{Chodos:1974je}
  A.~Chodos, R.~L.~Jaffe, K.~Johnson, C.~B.~Thorn and V.~F.~Weisskopf,
  Phys.\ Rev.\ D {\bf 9}, 3471 (1974).

\bibitem{Myhrer:1988ap}
  F.~Myhrer and A.~W.~Thomas,
  Phys.\ Rev.\ D {\bf 38}, 1633 (1988).

\bibitem{Young:2002cj}
  R.~D.~Young, D.~B.~Leinweber, A.~W.~Thomas and S.~V.~Wright,
  Phys.\ Rev.\ D {\bf 66}, 094507 (2002)
  [hep-lat/0205017].

\bibitem{Thomas:2008ga}
  A.~W.~Thomas,
  Phys.\ Rev.\ Lett.\  {\bf 101}, 102003 (2008)
  [arXiv:0803.2775 [hep-ph]].

\bibitem{Jaffe:1987sx}
  R.~L.~Jaffe,
  Phys.\ Lett.\ B {\bf 193}, 101 (1987).

\bibitem{Leinweber:2002qb}
  D.~B.~Leinweber,
  Phys.\ Rev.\  D {\bf 69}, 014005 (2004)
  [arXiv:hep-lat/0211017].


\bibitem{Wang:2007iw}
  P.~Wang, D.~B.~Leinweber, A.~W.~Thomas and R.~D.~Young,
  Phys.\ Rev.\ D {\bf 75}, 073012 (2007)
  [hep-ph/0701082].

\bibitem{Jenkins:1992pi}
  E.~E.~Jenkins, M.~E.~Luke, A.~V.~Manohar and M.~J.~Savage,
  Phys.\ Lett.\ B {\bf 302}, 482 (1993)
  [Phys.\ Lett.\ B {\bf 388}, 866 (1996)]
  [hep-ph/9212226].


\bibitem{Leinweber:2004tc}
  D.~B.~Leinweber {\it et al.},
  Phys.\ Rev.\ Lett.\  {\bf 94}, 212001 (2005)
  [arXiv:hep-lat/0406002].

\bibitem{Wang:1900ta}
  P.~Wang, D.~B.~Leinweber, A.~W.~Thomas and R.~D.~Young,
  Phys.\ Rev.\ C {\bf 79}, 065202 (2009)
  [arXiv:0807.0944 [hep-ph]].


\bibitem{Wang:2012hj}
  P.~Wang, D.~B.~Leinweber, A.~W.~Thomas and R.~D.~Young,
  Phys.\ Rev.\ D {\bf 86}, 094038 (2012)
  [arXiv:1210.5072 [hep-ph]].



\bibitem{Hogaasen}
H.~Hogaasen and F.~Myhrer, Phys.\ Rev.\ D {\bf 37}, 1950 (1988).

\bibitem{Gensini:1995ev}
  P.~M.~Gensini,
  [hep-ph/9512440].

\bibitem{Larin:1993tq}
  S.~A.~Larin,
  Phys.\ Lett.\ B {\bf 303}, 113 ,334 (1993)
  [hep-ph/9302240].

\bibitem{Wang:2013cfp}
  P.~Wang, D.~B.~Leinweber and A.~W.~Thomas,
  Phys.\ Rev.\ D {\bf 89}, no. 3, 033008 (2014)
  [arXiv:1312.3375 [hep-ph]].

\bibitem{Hall:2013dva}
  J.~M.~M.~Hall, D.~B.~Leinweber and R.~D.~Young,
  Phys.\ Rev.\ D {\bf 89}, no. 5, 054511 (2014)
  [arXiv:1312.5781 [hep-lat]].


\bibitem{Wang:2015sdp}
  P.~Wang, D.~B.~Leinweber and A.~W.~Thomas,
  Phys.\ Rev.\ D {\bf 92}, no. 3, 034508 (2015)
  [arXiv:1504.06392 [hep-ph]].


\bibitem{Leinweber:2003dg}
  D.~B.~Leinweber, A.~W.~Thomas and R.~D.~Young,
  Phys.\ Rev.\ Lett.\  {\bf 92}, 242002 (2004)
  [hep-lat/0302020].


\bibitem{Wang:2010hp}
  P.~Wang and A.~W.~Thomas,
  Phys.\ Rev.\ D {\bf 81}, 114015 (2010)
  [arXiv:1003.0957 [hep-ph]].

\bibitem{Allton:2005fb}
  C.~R.~Allton, W.~Armour, D.~B.~Leinweber, A.~W.~Thomas and R.~D.~Young,
  Phys.\ Lett.\ B {\bf 628}, 125 (2005)
  [hep-lat/0504022].

\bibitem{Armour:2008ke}
  W.~Armour, C.~R.~Allton, D.~B.~Leinweber, A.~W.~Thomas and R.~D.~Young,
  Nucl.\ Phys.\ A {\bf 840}, 97 (2010)
  [arXiv:0810.3432 [hep-lat]].

\bibitem{Hall:2013oga}
  J.~M.~M.~Hall, D.~B.~Leinweber and R.~D.~Young,
  Phys.\ Rev.\ D {\bf 88}, no. 1, 014504 (2013)
  [arXiv:1305.3984 [hep-lat]].




\bibitem{Young:2002ib}
  R.~D.~Young, D.~B.~Leinweber and A.~W.~Thomas,
  Prog.\ Part.\ Nucl.\ Phys.\  {\bf 50}, 399 (2003)
  [hep-lat/0212031].

\end{thebibliography}
\end{document}